# Free Will and Quantum Mechanics: Much Ado about Nothing

Stephen Boughn⁎

Departments of Physics and Astronomy, Haverford College, Haverford PA

In a recent series of papers and lectures [1-4], John Conway and Simon Kochen presented *The Free Will Theorem*. "It asserts, roughly, that if indeed we humans have free will, then elementary particles already have their own small share of this valuable commodity."[2] To be sure, perhaps the primary motivation of their papers was to place stringent constraints on quantum mechanical hidden variable theories[1], which they indeed do. Nevertheless, the notion of "free will" is crucial to the proof and they even speculate that the free will afforded to elementary particles is the ultimate explanation of our own free will. Conway and Kochen are eminent mathematicians with considerable knowledge of the formalism of quantum mechanics and I'm certainly not going to challenge the mathematics/logic of their proof. Rather, it's their premises with which I take issue. *Free will* and *determinism* are, for me, not nearly adequately clarified for them to form bases of a theoretical proof. In addition, they take for granted supplemental concepts in quantum mechanics that are in need of further explanation. Pragmatist that I am, it's also not clear to me what utility is afforded by "the free will theorem", i.e., what, if anything, follows from it. I suppose it would be useful if it were to help put an end to the endless philosophical discussions of free will and the persistent elaborations of hidden variable theories; however, that seems unlikely. Finally, despite the cheeky subtitle of my essay, I do think that the explicit introduction of free will into discussions of hidden variables and other interpretations of quantum mechanics would help expose foibles in many of those deliberations. For this reason, I consider the Conway-Kochen free will theorem to be a positive contribution to the philosophy of quantum mechanics.

---

⁎ sboughn@haverford.edu

[1] Hidden variable theories are proposals to provide deterministic explanations of quantum mechanics by the introduction of unobservable hypothetical entities. I've never found hidden variable theories to be particularly interesting or useful.

**Free Will and Determinism**

The concept of *free will* has been with us for millennia.[5] In short, free will indicates that under normal circumstances we have control of our actions. The immediate implication is that, unless we are physically forced or threatened, we are responsible for what we do. It's not difficult to imagine the utility of this notion. Theologians and jurists use free will to justify administering retribution to miscreants and criminals. For how could one otherwise reward or punish someone who is not in control of their actions? The problem came when philosophers and theologians wrestled with the problem of a god or gods who determine everything that happens. If God is responsible for all that happens, how can individuals possibly be held responsible for anything they do? From the time of Plato to the present, philosophers and theologians have tied themselves in knots attempting to resolve this conundrum.

In the modern era, God has taken a back seat to physical determinism. After Galileo, Newton, and Laplace, it was thought that, according to the laws of nature, all physical phenomena are uniquely determined by the events that precede them. Because most scientists consider human thought to be a consequence of physical activity in the brain, the conflict of free will and determinism continued to present a philosophical problem and so the dilemma of how to embrace both free will and determinism continued. To be sure, the ancient Greek atomists had a mechanistic view of nature wherein all objects, including humans, are composed of atoms that move according to laws of nature, i.e., no gods involved. Epicurus was able to preserve free will by postulating that atoms undergo random, minute swerves that supersede the laws of nature.[5] In modern times it is quantum mechanics that provides the same sort of randomness that frustrates Newtonian determinism. On the other hand it's not at all obvious how randomness can result in free will, so the philosophical conundrum continues.

**The Free Will Theorem**

The following is only a thumbnail sketch of the Conway-Kochen theorem, but it will suffice in providing a backdrop for my quandary about the premises of their proof.

Conway and Kochen utilize a clever gedanken experiment that involves two spin 1 particles. After being prepared in a joint singlet state, the two particles move apart, one remains on Earth and the other travels to Mars.[1] An observer on Earth measures the squares of the orthogonal *x,y,z* components of the spin, $S_x^2, S_y^2, S_z^2$, of one particle.[2] According to quantum mechanics (and confirmed by experiment) the observed values of $S_x^2, S_y^2$, and $S_z^2$ will always be 1, 0, 1 in some order but before the measurement one does not know which particular order will be observed. This follows from the usual probabilistic nature of quantum mechanics. An observer on Mars measures the square of the other particle's spin component in the *w* direction, $S_w^2$. Because the particles were prepared in a singlet state, the value of $S_w^2$ will be the same as $S_x^2$ if *w* is in the *x* direction, the same as $S_y^2$ if *w* is in the *y* direction, and the same as $S_z^2$ if *w* is in the *z* direction.

Now assume the two observers possess "free will", that is, they are free to choose the *x,y,z,* and *w* directions independent of each other and of the past history of the universe. If the two sets of measurements are performed within less than the light travel time between Earth and Mars, the Earth observer's choice of the orthogonal *x,y,z* directions cannot depend on the Mars observer's choice of the *w* direction and vice versa. Neither can the result of the spin measurement of either observer depend on the result obtained by the other. Such dependences would constitute clear violations of special relativity. Suppose that, contrary to the free will theorem, neither elementary particle has free will. That is, the results of the spin measurements depend only on the choices of the axes and the past histories of the particles. Conway and Kochen express this as: the response of either measurement is "a function of properties of that part of the universe that is earlier than this response with respect to any given inertial frame."[2]

They restrict the *w* direction to be one of 33 special directions (the Peres configuration) and the *x,y,z* directions to be one of the 40 orthogonal triplets formed out of these 33 directions. By pairing the *w* directions with various *x,y,z* directions, Conway and Kochen demonstrate that the result of the Earth observer's measurement is a 101 function, i.e., "a function defined on the sphere of possible directions taking each orthogonal triple to 1, 0, 1 in some order."[2] The arguments of this function are *x,y,z* in addition to parameters specifying the properties of the past history of that particle including any statistical properties that represent hidden variables. The problem is that a previous theorem by Kochen and Specker [6] demonstrated the nonexistence of such a 101 function with the implication that the response of the particle is "not a function defined on the sphere of possible directions taking each orthogonal triple to 1, 0, 1 in some order." Therefore, the particle's response cannot be determined prior to its measurement; to wit, like the observer, the particle also possesses at least a modicum of free will. The proof and implications of the free will theorem involve subtleties that the authors treat in detail. In particular, they convincingly argue "…there can be no correct relativistic deterministic theory of nature. In particular, no relativistic version of a hidden variable theory…can exist." Moreover, the free will theorem "has the stronger implication that there can be no relativistic theory that provides a mechanism for reduction [collapse of the wave function]."[2] Now, on to my questions about the premises assumed in the free will theorem.

---

[2] The quantum mechanical operators, $S_x^2, S_y^2$, and $S_z^2$, commute with each other and so the spins in these directions can be measured simultaneously.

**The Meanings of Free Will and Determinism**

The notions of *free will* and *determinism* in the above account were taken to be more or less self-evident; however, let's probe a bit deeper into their meanings. So just what is meant by "free will"? Most might respond that free will means that you can do anything you *will* to do. Of course there are restrictions: one cannot (should not) do what is prohibited by law or forbidden by a religion or that would offend one's family, friends and community. Even so, one might decide to act independent of these considerations. Certainly one's decision to act depends on a plethora of outside factors, including the weather. Now I've seemed to equate free will with the ability to make decisions but that's also not quite right. What is it about a decision that renders it the consequence of free will? Perhaps, for a decision or an action to be truly free it must be unpredictable because if what you're going to do can be predicted, how could your action be truly free. On the other hand, the ability to predict the actions of others is an absolutely necessity for a well functioning society. In fact, we believe that the more information we have about people, the better we will be able to predict their actions. If the only time one's actions are truly free is when they're unpredictable then free will is a truly frightening phenomenon.

Conway and Kochen argue that they require "…only a minuscule amount of human free will", namely the choice of the axes relative to which one measures particle spin. However, they also point out that

> It is hard to take science seriously in a universe that in fact controls all the choices experimenters think they make. Nature could be in an insidious conspiracy to "confirm" laws by denying us the freedom to make the tests that would refute them. Physical induction, the primary tool of science, disappears if we are denied access to random samples. It is also hard to take seriously the arguments of those who according to their own beliefs are deterministic automata! [1]

I think we would all agree with this statement; however, it's not clear that a shrewd judge of human character would not be able to predict the choices of a particular experimenter. In their statement of the free will theorem, they explain: "It is usually tacitly assumed that experimenters have sufficient free will to choose the settings of their apparatus in a way that is not determined by past history."[1] But what do they mean by "determined by past history"? This brings us to the notion of determinism, the antithesis of free will.

In their proof, Conway and Kochen give an explicit mathematical example of "determined by past history", namely the 101 function that determines the outcomes of the three spin measurements performed by the Earthbound observer. They demonstrate that such a function takes the form of $F_{101}(x, y, z, \alpha)$ where $x, y, z$ represents the experimenter's free choice of the measurement axes and $\alpha$ represents the properties of that part of the universe at times prior to the measurement. So they mathematicized "past history" as a function of the properties of the universe in the past. But what are these properties and how might we identify them? Conway and Kochen define a *property* as information, encoded in a single bit, as to whether or not a particular *event* has occurred. Even if I knew precisely what is meant by an event, which I don't, I have no idea how one might gather information about all conceivable events. Even if we lived in a classical universe, which we do not, I suppose this might mean all possible positions and momenta of all $10^{80}$ particles in the known universe.[3] Even then I'm at a total loss for how one might encode this information in a function that nominally would predict both the decision of the experimenter and the results of the spin measurements. In principle, this would surely entail a super theory of everything, one that includes not only the 4 forces of nature but also the complexity required by chemistry, biology, psychology, economics, history, etc. It would have to be an exact super theory of everything and not the patchwork of theories we now have because of discrepancies of these patches in their regions of overlap. No such theory exists (nor probably ever will) and certainly no human and I dare say no computer, quantum or otherwise, could generate such a function, even in principle.

---

[3] I've estimated elsewhere [7] that the number of such events is on the order of $10^{10^{82}}$!

Okay, I realize that Conway and Kochen don't consider the 101 function to be a human construction but rather as simply following from the *laws of nature*, whether or not we know these laws. But now we must ask: What are the laws of nature and why do we assume that they are mathematical laws? The concept of *law* is again a human construct. Laws are proclaimed by kings and governments in order to maintain stable societies. To be sure, as far back as Plato, philosophers acknowledged laws that were conferred by God or nature. Many physicists are realists to the extent that they subscribe to the belief that the natural world evolves according to nature laws; although, there is certainly no direct evidence that this is the case, no more than there is that the world evolves according to God's laws. So what are these laws of nature? Certainly Newton's law of gravitation would be offered as one such law (perhaps updated to Einstein's general relativity), while Maxwell's classical electrodynamics would be another. Now both of these laws are represented by mathematical equations so it naturally follows that nature evolves according to mathematical laws, which then might lead to Conway and Kochen's mathematical expression of determinism. But in the end, Newton's and Maxwell's equations are human creations, mathematical models and, therefore, so also are mathematical expressions of determinism. As such, these concepts only have validity to the extent that their uses are clearly specified and, as noted above, there is certainly nowhere near enough specificity to claim that the past history of the world can be expressed as an explicit function, e.g., $F_{101}(x, y, z, \alpha)$.

**Clarifying Other Concepts**

There are several other concepts that the authors take to be more or less self-evident. One of these is *causality*, which they consider to be on the same level as a law of nature. They note [2]:

> It is the experimenters' free will that allows the free and independent choices of *x, y, z*, and *w*. But in one inertial frame—call it the "A-first" frame—B's experiment will only happen some time later than A's, and so *a*'s response cannot, by temporal causality, be affected by B's later choice of *w*.

Kochen points out [3], "Causality is not a consequence of the formalism in either classical or quantum physics, but is so imbedded in our thinking that it is generally accepted as a universal principle." Maybe so, but causality is another concept that needs further clarification. So just what do we mean by causality? Like "free will", the notion of "cause" is often used in everyday language. When things happen, it's human nature to attribute these events to causes. But in the context of the free will theorem we would like something more definitive. In Newtonian physics, cause is directly related to determinism. For example, if a projectile of mass $m_1$ and velocity $v_1$ strikes and sticks to a stationary mass $m_2$, the two will move off together with a velocity $\frac{m_1}{m_1+m_2} v_1$. In this case, its interaction with the projectile constitutes the *cause* of the subsequent motion of the initially stationary mass. But this example of causality is tied to a particular formalism, Newtonian mechanics. It is not the law of nature[4] invoked in the free will theorem as required for the deterministic expression of the results of spin measurements. So like *free will* and *determinism*, a precise definition of *causality* is absent[5] and so causality is inappropriately invoked in the proof of the free will theorem.

As indicated above, another consequence of the free will theorem is that no relativistic theory can provide a physical mechanism for collapse of the wave function (quantum state reduction). "A future theory may reasonably be expected to describe more fully exactly which 'textures' [interactions] will cause reductions, but the Free Will Theorem shows that no such theory will correctly predict the results of these reductions."[1] *Collapse of the wave function* was introduced by von Neumann and Dirac in the early days of quantum mechanics when physicists were trying to understand the role of measurements in quantum mechanics and, in general, to make sense out of mysterious aspects of the quantum formalism. They treated wave function collapse as a physical process that was in need of explanation. But wave functions are simply constructs of the mathematical models of quantum phenomena and as such are human inventions that derive their meaning from the way physicists use them. They are patently not physical phenomena that require an ontological account. Physicists assign a wave function to a particular system based on what is known about the system and then use the wave function to predict the evolution of the system and predict the results of measurements. If and when more information is available, one simply assigns a new wave function to the system. Wave functions are not unique physical properties of systems nor wave function collapse a physical process. Conway and Kochen certainly do not agree with this characterization. They maintain [1], "We believe that the reduction is a real effect that will only be explained by a future physics, but that current experiments are already informative." If so, what is the evidence that "reduction is a real effect"? None that I know of.

---

[4] While causality may not be a law of nature, there is no doubt that the concept has been extremely useful in conjuring new models of nature. However, the source of scientific creativity is well beyond my understanding and I'll refrain from addressing this topic.

[5] According to philosopher Bas van Fraassen [9], "In some cases…a causal theory will achieve… empirical adequacy, in other cases it will not, and that is just the way the world is. The causal terminology is descriptive, in any case, not of the … phenomena, but of the proffered theoretical models. So pervasive has been the success of causal models in the past, especially in a rather schematic way at a folk-scientific level, that a mythical picture of causal processes got a grip on our imagination."

      Kochen [3] does point out subtleties in the notion of a quantum state and attributes the source of claims of quantum non-locality to the "conflation of the two concepts of state and property." He restricts "state" to be the quantum state (wave function) of a system with the usual implications for the statistical outcome of a future measurement and "property" to be the outcome of that measurement. I'm sympathetic with this characterization; however, it is evident that Conway and Kochen consider both "state" and "property" to be physical aspects of a system. I don't consider either to be so. I've already considered the status of "state", i.e., wave function. But also "property", i.e., result of a measurement, is simply our description of an observed event. Even the notion of "measurement" is not a part of quantum mechanics but something, like free will, that lies completely outside the formal theory. Consider Bohr's description of a

measurement:

> The decisive point is to recognize that the description of the experimental arrangement and the recordings of observations must be given in plain language, suitably refined by the usual terminology. This is a simple logical demand, since by the word 'experiment' we can only mean a procedure regarding which we are able to communicate to others what we have done and what we have learnt. [8]

Nothing in this description refers to a physical theory, quantum or classical. Conway and Kochen would undoubtedly support such an account of measurement. They describe the spin experiment as a combination of the observer's *choice of the orientation* of the apparatus and the *spot on the screen* made by the particle, and then make the point that neither of these two aspects of the measurement is a theoretical concept and so are independent of any particular theory. "Our dismissal of hidden variable theories is therefore much stronger than the previous ones that presuppose quantum mechanics."[1] However, if this is the case it's difficult to understand, in the absence of a formal model, how a mathematical expression $F_{101}(x, y, z, \alpha)$ can be assigned to the spot on the screen as required in the free will theorem.

**Final Remarks**

With regard to this essay, I might be accused of expounding my personal views of the philosophical foundations of physics [10] rather than finding fault with the Free Will Theorem. Guilty as charged! However, any time I come across an analysis of the foundations of quantum theory, especially if that examination employs ill-defined concepts like wave function collapse, quantum measurement, and non-locality or seemingly self-evident notions like determinism, causality and free will, I feel compelled to respond. In the opening paragraph I remarked that the Conway/Kochen theorem constitutes a positive contribution to the philosophy of quantum mechanics. Why? Their work directly confronts those aspects of science that lie outside the formalism of quantum theory. These include free will, measurement, and causality. Other such analyses often obfuscate this distinction. In the authors' words describing other disproofs of hidden variable theories[1], "They involve questionable notions such as 'elements of reality', counterfactual conditionals, and the resulting unphysical kinds of locality." By bringing

10the notion of "free will" to the forefront, even including it in the titles of their papers, Conway and Kochen force us to directly confront those sociological aspects of physical interpretation that are necessarily removed from physical theory. They directly point to this in a closing remark of their first paper. A great advantage of the free will theorem "is that it applies directly to the real world rather than just to theories".

So what do I think about the conclusion of the free will theorem; do elementary particles possess free will? My answer is a qualified "yes" with the caveat that the notion of free will is certainly not well defined and has little utility beyond its applications in jurisprudence and religion.

## Acknowledgements

I'm grateful to many friends who have tolerated my sojourns into philosophy over the years. They include Bob, Dave, Ed, Eliot, Freeman, Hal, Hojung, Jeff, Jim, Juan, Larry, Lyman, Marcel, Mike, Norm, Peter, Shannon, Susan,Tony,… (First names only; I wouldn't want anyone's intellect to be impugned. ☺)